\documentstyle[12pt]{article}
\input epsf

\newcommand{\beq}{\begin{equation}}
\newcommand{\eeq}{\end{equation}}

\begin{document}
\renewcommand{\thefootnote}{\fnsymbol{footnote}}

\vspace{3cm}
\begin{center} \Large
{\bf ITEP Lectures in Particle Physics} 
\end{center}
\vspace*{.3cm}
\begin{center} {\Large
Mikhail Shifman } \\
\vspace{0.4cm}
{\it  Theoretical Physics Institute, University of Minnesota,
\\
Minneapolis, MN 55455, USA}
\end{center}


\section{Foreword}

My career in theoretical high energy physics began 25 years ago.
I was lucky -- its beginning coincided with a very exciting time
in this field, when a major breakthrough in our understanding of 
nature took 
place. The Standard Model of fundamental interactions was born,
and Quantum Chromodynamics (QCD) gradually emerged as {\em the} 
theory 
of  hadronic matter. Unlike many other theories created later 
whose 
relevance to nature is still a big question mark, these two will 
definitely stay 
with us forever. And, fortunately, I happened to be in the right place 
at the right 
time  so that I could appreciate these discoveries 
early.
The infancy of any true theory creates great opportunities for young 
researchers who suddenly find themselves pioneers in {\em terra 
incognita},
with so many interesting, important and challenging problems 
around.

For about twenty years, I was a member of the ITEP 
 theory group\footnote{For those who do not know: ITEP is an 
abbreviation for Institute of Theoretical and Experimental Physics in 
Moscow.}. 
During 
these years I gave many lectures at different schools of physics. 
Some of them 
are presented below.  As a matter 
of fact, some of the lectures are recent, from the period after I  
left ITEP. Still, I consider them ITEP
lectures since I learned many of the ideas in them from my formal 
and 
informal teachers at ITEP. I am especially grateful to B. Ioffe, A. 
Vainshtein 
and V. Zakharov. 

Since the ITEP theory group of the seventies and the eighties was 
quite an 
endemic phenomenon,  it 
seems 
worthwhile to introduce it to the reader before proceeding to the 
scientific issues. 

\subsection{Glimpses of ITEP}

ITEP was more than an institute. It was our refuge where the 
insanity 
of the 
surrounding  reality was, if not eliminated,  was 
reduced to a bearable level. Doing physics there was something 
which gave a meaning to our lives, making it interesting and even 
happy.
Our theory group was like a large family. As in any family, of 
course, this 
did not mean that everybody loved everybody else, but we knew 
that we had 
to stay together and to rely on each other, no matter what, in order 
to survive
and to be able to continue doing physics. This was considered by our 
teachers 
to be the most important thing, and this message was always being
conveyed, in more than one way, to young people joining the group. 
We had a 
wonderful feeling of stability in our small brotherhood.  A feeling so 
rare in the
Western laboratories where  whirlpool of 
postdocs,
visitors, sabbatical years come and go, there are a lot of new faces, 
and a lot of people 
whom you 
do not care so much about.

The rules of survival were quite strict. First, seminars -- what is now 
known 
worldwide as the  famous Russian-style seminars. The 
primary goal of the speaker was to {\em explain} to the audience his 
or her 
results, not merely to advertise them. And if the results were 
non-trivial, or 
 questionable or just unclear points would surface in the course of 
the 
seminar, 
the standard two hours were not enough to wind up. Then the 
seminar 
could last 
for three or even four hours, until either everything was  clear or 
complete
exhaustion, whichever came first. I remember one seminar in 
Leningrad  
in 1979, when Gribov was still there, which started at eleven in the 
morning. A lunch break was announced from two  to three, and then 
it 
continued 
from three till seven in the evening. In ITEP we had at least three 
theoretical 
seminars a week, a formal one on Mondays, and an informal coffee 
seminar 
which at first took place every Friday at 5 o'clock, when the official 
work day 
was over, but later was shifted to Thursdays, the same time. Usually, 
these 
were by far, the most exciting events of the week. The leaders and 
the 
secretaries of the seminars were supposed to find exciting topics, 
either by 
recruiting ITEP or other ``domestic" authors, or, quite often, by 
picking up 
a paper or a preprint from the outside world and asking somebody to 
learn 
and  report the work to the general audience. This duty was 
considered to 
be a moral obligation. The tradition dated back to the times when 
Pomeranchuk was the head of the theory group, and its isolation had 
been even  more severe than in my times. As a matter of 
fact, in
those 
days there were no preprints, and getting fresh issues of Physical 
Review or 
Nuclear Physics was not taken for granted at all. When I, as a 
student, joined 
the group -- this was a few years after Pomeranchuk's death -- I was 
taken, 
with pride, to the Pomeranchuk memorial library, his former office 
where a 
collection of his books and  journals  was kept. 
Every paper, in every issue was marked by Chuk's hand (that's how 
his 
students and colleagues 
would refer to him), either with a minus or a plus sign.
If 
there was plus, there was a name of one of his students who had 
been asked 
to dig into the paper and give a talk for everyone's benefit. This 
was not the end of the story, however. Before the scheduled day of 
the 
seminar Pomeranchuk would summon the speaker-to-be to his office 
to give a 
pre-talk, to him alone, so that he could judge whether the 
subject 
had been worked out with sufficient depth and that the speaker was 
ripe 
enough to 
face the general audience and its bloodthirsty questions. In my 
times, 
the 
secretaries of the seminars were less inclined to sacrifice themselves 
to that 
extent, but, still, it was not uncommon that a pre-talk would be 
arranged
for an unknown, young or inexperienced speaker.

Scientific reports of the few chosen  to travel abroad 
for a 
conference or just to collaborate for a while with Western physicists, 
were
an unquestionable element of the seminar routine. Attending an 
international 
conference by A or B by no means was considered as a personal 
matter of A 
and B. Rather, these rare lucky guys were believed to be our 
ambassadors,
and were supposed to represent the whole group. In 
practical terms,
this meant that once you made your way to a conference,  
you could be asked to present important results of other members of 
the 
group. Moreover, you
were supposed to attend as many talks as physically possible, 
including those 
which did not exactly belong to your field, make extensive notes and 
then, 
after returning,
deliver  an exhaustive report of all new developments discussed,  all 
interesting questions raised, rumors, etc. The scientific rumors, as 
well as 
non-scientific impressions were like an exotic dessert, usually served 
after 
nine. I remember that, after his first visit to the Netherlands,  
Simonov 
mentioned  that he was very surprised to see a lot of people on the 
streets 
just smiling. He said he could not understand why they looked so 
relaxed.
And then he added that he finally figured out why: ``... because they 
were not 
concerned with building communism..." This remark almost 
immediately 
became known to ``Big Brother" who was obviously watching  us this 
evening, as usual, and it cost Simonov a few years of sudden 
``unexplainable 
allergy" to any Western exposure. His health condition, of 
course, 
would 
not allow him to accept any invitation to travel there. I can not help 
mentioning another curious episode with Big Brother. Coffee,  which 
we used to 
have during the coffee seminars was being prepared in turn, by all 
members 
of the group. Once, when it was Ioffe's turn, he brought a small
bottle of cognac and  added a droplet or two in every cup. I 
do not 
remember why, perhaps, it was his birthday or something like that. 
This was Friday evening, and very early next  Monday morning  he 
was summoned to the corresponding ITEP branch office 
to give 
explanations concerning his ``obviously subversive activities".

The coffee seminars typically lasted till nine, but sometimes much 
later, for 
instance, in the stormy days of the November revolution in 1974. 
The few 
months following the discovery of $J/\psi$ were the star days of QCD 
and, 
probably, the highest emotional peak of the ITEP theory group. 
Never were the 
mysteries of 
physics  taken so close to our hearts as then. There was a
spontaneously arranged team of enthusiasts working practically in a 
non-stop 
regime. A limit to our discussions was set only by the schedule of the 
Moscow 
metro -- those who needed to catch the last train had to be leaving 
before 1 
a.m. 

The ITEP seminars were certainly one of the key 
elements 
 in shaping the principles and ideals of our small 
community, but 
not 
the only one. The process of selecting  students  who 
could eventually grow up into 
particle theorists played a crucial role and was, probably, as 
elaborate as
the process of becoming a knight of the British crown. Every year we 
had 
about 20 new students at the level roughly corresponding to that of
the graduate students in  American universities. Mostly, they  came 
from  the Moscow Institute for Physics and Technology, a small elite 
institution 
near the city, a counterpart of MIT in the States. Some students were 
from the
Moscow Engineering and Physics Institute, and a few from the 
Moscow 
State 
University. They were offered (actually, obliged to take) such a 
spectrum of
courses in special disciplines which I have never heard of anywhere 
else in 
the world: everything  from radiophysics and accelerator 
physics; 
several levels of topics in quantum mechanics, including intricacies of 
theory 
of scattering; radiation theory and nuclear physics; mathematical 
physics (consisting of several separate parts);  not 
less than three courses in particle phenomenology (weak, 
electromagnetic and 
strong interactions); quantum electrodynamics, numerous 
problem-solving 
sessions, etc.  And yet, only those who successfully passed 
additional 
examinations, covering  the famous course of theoretical physics by 
Landau 
and Lifshitz,  were allowed, after showing  broad erudition and 
ingenuity in 
solving all sorts of tricky
problems, to join the theory group. Others were supposed to end 
up as experimentalists or engineers. Needless to say that the process 
of 
passing these examinations could take months, and even years, and 
was 
notoriously exhausting, but  there never was lack of volunteers to try 
their 
luck. They were always seen around Ter-Martirosian and Okun 
who were 
sort of responsible for the program. It should be added that the set 
of values 
to be passed from the elders to the young generations included the 
idea that 
high energy physics is an experimental science that {\em must}  be 
very 
closely 
related to phenomena taking place in nature. Only those theoretical 
ideas  which, at the end of the day,  could produce  a number 
which could 
be confronted with phenomenology were 
cherished.  Too abstract and speculative 
constructions, and theoretical phantoms,  were not encouraged, to put 
it 
mildly.  
The atmosphere was strongly polarized against what is now 
sometimes called
``theoretical theory".  Even extremely bright students,  who were too 
mathematically oriented, like, say, Vadim Knizhnik, were having 
problems 
in passing these examinations. Vadim, by the way, never made it to 
the end,
got upset and left ITEP. Well, nothing is perfect in this world, and I 
do not
want to make an impression that the examination routine in the ITEP 
theory 
group was without flaws. 

The ITEP theory group was large -- about 50 theorists -- and 
diverse. 
Moreover, it was a natural center of attraction for the whole Moscow 
particle 
physics 
community. Living in the capital of the last world empire had its 
advantages. 
There is no question, it was the evil empire but, what was good, as it 
usually 
happens 
with any empire, all intellectual forces tended to cluster in the 
capital. So, we 
had a very dynamic group where virtually every direction was 
represented 
by at least several theorists, experts in the given field. If you needed 
to learn 
something new there was an easy way to do it, much faster and more 
efficiently than through reading journals or textbooks. You just 
needed to talk 
to the right person. Educating others, sharing your knowledge and 
expertise 
with
everybody who might be interested was another rule of survival in 
our 
isolated community. In such an environment, different discussion 
groups and 
large collaborations were naturally emerging all the time, creating a 
strong 
and positive coherent effect. The brain-storming sessions used to 
produce,
among other results, a lot of noise, so once you were inside the old 
mansion
occupied by the theorists, it was very easy to figure out which task 
force was 
where -- just step out in the corridor and listen. And, certainly, all 
these 
sessions were open to everybody. 

Now I would like to mention one more aspect which concerns me at 
present, 
a very strong pressure existing in our  community,  to 
stay in the 
``mainstream", to work only on fashionable directions and problems
which,  currently,   are under investigation in 
dozens of other laboratories . This pressure is especially damaging 
for young people who  have little alternative. Of course, a certain 
amount of 
cohesion is needed, but the scale of the 
phenomenon we are witnessing now is unhealthy, beyond any doubt.
The isolation of the ITEP theory group had a positive side effect.
Everybody, including the youngest members, could afford
working on problems not belonging to the fashion of the day, without 
publishing a single line for a year or two. Who cared about what we 
were doing there anyway? This was okay. On the other hand, it was 
considered indecent to publish    results of 
dubious novelty,
incomplete results (of the status report type) or just  papers with too 
many words per given number of formulae. Producing   dense papers 
was a norm.  This style, which was probably perceived by the 
outside readers as a chain of riddles, 
is partly explained by  tradition, presumably dating back to 
the 
Landau times. It  was also due to specific Soviet conditions, where
everything was regulated, including the maximal number
of pages any given paper could have.  Compressing  
derivations and arguments to the level considered acceptable, was an  
art
which had its grandmasters. Arkady Vainshtein was especially good 
at inventing all sorts of tricks allowing him to squeeze in extra 
formulae 
with very few explanatory remarks. I remember that in 1976, when 
we were 
working on 
the  large JETP paper on   penguins in  weak decays\footnote{By 
``we"
I mean Zakharov, Vainshtein and myself. Arkady Vainshtein had a 
permanent position at the Budker Institute of Nuclear Physics in 
Novosibirk. He commuted between Moscow and Novosibirsk for 
many years.}
 we had to make  30 pages out of the original 60-page
preprint version, and
he 
managed to do that
losing no equations and even inserting a few extra! This left a 
strong impression on me. 

By the way, about penguins. From time to time students ask  
about how this word could possibly penetrate high energy physics. 
This is a 
funny story, indeed.
The first paper where the graphs that are now called penguins were 
considered in the weak decays appeared in JETP Letters in 1975, and 
there they did
not look like penguins at all. Later on they were made look line 
penguins 

\vspace{0.5cm}
\epsfxsize=25truemm
\centerline{\epsfbox{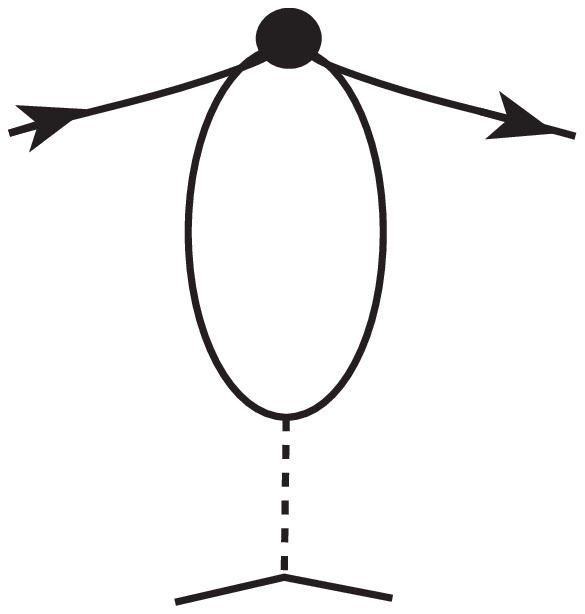}}

\vspace{0.5cm}
\noindent
and called penguins by John Ellis. Here is his story as
he recollects it himself.

\vspace{0.3cm}

    ``Mary K. [Gaillard], Dimitri [Nanopoulos] and I first got interested 
in what are now called
penguin diagrams while we were studying CP violation in the 
Standard Model in 1976... The penguin name came in 1977, as 
follows.

    In the spring of 1977, Mike Chanowitz, Mary K and I wrote a 
paper on GUTs predicting the $b$  quark mass before it was found. 
When it was found a few weeks later, Mary K, Dimitri, Serge Rudaz 
and I immediately started working on its phenomenology. That 
summer, there was a student at CERN, Melissa Franklin who is now 
an experimentalist at Harvard. One evening, she, I and Serge went to 
a pub, and she and I started a game of darts. We made a bet that if I 
lost I had to put the word penguin into my next paper.
She actually left the darts game before the end, and was replaced by 
Serge, who beat me. Nevertheless, I felt obligated to carry out the 
conditions of the bet.

    For some time, it was not clear to me how to get the word into this 
$b$ quark paper that we were writing at the time. Then, one 
evening, after working at CERN, I stopped on my way back to my 
apartment to visit some friends living in Meyrin where I smoked 
some illegal substance. Later, when I got 
back to my apartment and continued working on our paper, I had a 
sudden flash that the famous diagrams look like penguins. So we put 
the name into our paper, and the rest, as they say, is history."

\subsection{About these lectures} 

Below the reader will find  lectures devoted to different topics
in theoretical high-energy physics 
which occupied me during the last 15 years.  The 
choice of topics might seem   somewhat chaotic, at first sight. The 
selection criteria were simple: I tried to pick up only those topics 
which are of interest today. Besides, I was limited by 
 the fact that 
not all lecture notes are available; some have never been published 
and have disappeared with time. 

These are {\em lectures}, not reviews; they were intended for 
beginners
-- mostly graduate students -- who were just about to submerge
into the subject and needed some initial impetus and general idea 
and  guidance. Therefore, the pedagogical element was most 
important. I made no attempt at complete coverage,  the lists of 
references are usually quite fragmentary and so on. These lectures 
can be used for the initial exposure. Those who would like to master 
the corresponding topics in full, will need to proceed to more detailed 
reviews and the original literature. At the end of each lecture I 
recommend a few
sources for further studies. 

The lectures written in the eighties were revised and updated 
specifically for this Volume. I tried to keep
the revisions minimal, eliminating only the most evident drawbacks 
and resisting  the temptation to completely 
rewrite them. This process would take too much time and
the Volume would never appear had I not settled on a compromise.
I had a special reason to hurry up: at the end of October 1995 ITEP 
celebrates its 50 anniversary, and I made up my mind to complete 
the work by then. 

Here is the list of the lectures which will be presented in this 
Volume:

1). Lectures on Heavy Quarks in Quantum Chromodynamics

2).  Beginning Supersymmetry (Supersymmetry in Quantum 
Mechanics)

3). ABC of Instantons (V. Novikov, A. Vainshtein, V. Zakharov + M.S.)

4). Instantons at High Energies

5). Instantons Versus SUSY (A. Vainshtein, V. Zakharov + M.S.)

6). Miracles of Supersymmetric Gauge Dynamics

7). Two-Dimensional Conformal Field Theory: A Primer

8). New Findings in Quantum Mechanics

\end{document}